*Article*

# Dialectical GAN for SAR Image Translation: From Sentinel-1 to TerraSAR-X

**Dongyang Ao [1,2], Corneliu Octavian Dumitru[1], Gottfried Schwarz[1] and Mihai Datcu [1,\*]**

1. German Aerospace Center (DLR), Münchener Str. 20, 82234 Wessling, Germany; corneliu.dumitru@dlr.de gottfried.schwarz@dlr.de
2. School of Information and Electronics, Beijing Institute of Technology, Beijing 100081, China; aodongyang@foxmail.com (D.A.);
* Correspondence: mihai.datcu@dlr.de; Tel.: +49-815-328-1388

**Abstract:** Contrary to optical images, Synthetic Aperture Radar (SAR) images are in different electromagnetic spectrum where the human visual system is not accustomed to. Thus, with more and more SAR applications, the demand for enhanced high-quality SAR images has increased considerably. However, high-quality SAR images entail high costs due to the limitations of current SAR devices and their image processing resources. To improve the quality of SAR images and to reduce the costs of their generation, we propose a Dialectical Generative Adversarial Network (Dialectical GAN) to generate high-quality SAR images. This method is based on the analysis of hierarchical SAR information and the "dialectical" structure of GAN frameworks. As a demonstration, a typical example will be shown where a low-resolution SAR image (*e.g.,* a Sentinel-1 image) with large ground coverage is translated into a high-resolution SAR image (*e.g.,* a TerraSAR-X image). Three traditional algorithms are compared, and a new algorithm is proposed based on a network framework by combining conditional WGAN-GP (Wasserstein Generative Adversarial Network - Gradient Penalty) loss functions and Spatial Gram matrices under the rule of dialectics. Experimental results show that the SAR image translation works very well when we compare the results of our proposed method with the selected traditional methods.

## 1. Introduction

In remote sensing, SAR images are well-known for their all-time and all-weather capabilities. In the 1950s, the first SAR system was invented [1]. However, the design and implementation of a SAR system is a complex system engineering and costs many resources, both in money and intellectual effort. Therefore, most SAR instruments on satellites are supported by government organizations. For example, the German Aerospace Center (DLR) and EADS Astrium had jointly launched TerraSAR-X in 2007 [2] and TanDEM-X in 2010 [3]. The Canadian Space Agency (CSA) had launched in 1995 the RADARSAT-1 and in 2007 the RADARSAT-2 satellites [4], while the Italian Ministry of Research and the Ministry of Defence together with the Italian Space Agency (ASI) had launched the COSMO-SkyMed -1, 2, 3, and 4 satellites in 2007, 2008 and 2010 [5]. The European Space Agency (ESA) had launched the Sentinel-1 SAR satellite in 2014 [6]. In addition, there are many governments and institutions having launched their own SAR satellites [7], [8]. Nowadays, SAR has become one of the most valuable tools for remote sensing of the Earth and its environment.

In the era of big data, deep learning can accommodate large amount of data and generate promising new applications. With the recent development of deep learning, image translation is an easy way to obtain high-quality SAR images. "Translation" is a word borrowed from the linguistic field which denotes the change from one language to another one. This translation is often applied when one language is hard to understand while another one is more familiar to us. Though the two languages have different vocabularies and grammars, the translation is premised on the identity of the contents. In general for image translation there are two "sides" of the translation, namely the two images coming from different sensors. In this paper, we demonstrate a typical example where a low-resolution SAR image (*e.g.,* a Sentinel-1 image) with large ground coverage is translated using deep learning into a high-resolution SAR image (*e.g.,* a TerraSAR-X image). To some extent, this kind of translation is related to super-resolution and neural style transfer.

From 2013, deep learning has becomes a popular tool for many applications, such as image recognition, classification, semantic segmentation, target detection, etc. The first milestone in deep learning based image translation is Gatys *et al.*'s paper [9]. They introduced the Visual Geometry Group (VGG) networks, a pre-trained neural network used for ImageNet in order to define the content and "style" information of images, which provides a framework for image translation under the background of deep learning. Within a neural network based framework, many researchers have proposed their own methods for their specific purposes [10], [11]. The second milestone is the invention of Generative Adversarial Networks (GANs) that was made by Goodfellow *et al.* [12]. As a generative neural network, it seems that a GAN is well-suited for image translation. According to the conception presented in [13], the image translation can be regarded as the *"pix2pix"* task, and the authors of [13] have unitized a conditional GAN to carry out image translations. Inspired by this paper, we think that we can apply these algorithms to do SAR image translation. In SAR image processing, there are many papers about how to use deep learning for classification, segmentation, etc. [14], [15]. However, little attention has been paid to the translation between different SAR instruments using deep learning.

Translation of Sentinel-1 data to high-resolution images like TerraSAR-X has attracted great interest within the remote sensing community. First, the high resolution of TerraSAR-X generates SAR images rich in information that allow innovative applications. Second, the wide area coverage of Sentinel-1 images reduces the need for multiple acquisitions and decreases the demand for high-cost data. Third, it is much easier for researchers to access Sentinel-1 images than TerraSAR-X images because the Sentinel-1 images are freely available, while the TerraSAR-X images are usually commercial. To meet these requirements for high-quality data, we propose a "Dialectical GAN" method based on the analysis of the hierarchical SAR information and the "dialectical" structure of GAN frameworks. The data used for validation is covering urban areas, so we can apply a spatial matrix to extract geometrical arrangement information. By using a GAN, we were able to achieve good results with fine visual effects and our indicators show that our proposed method is better than the existing traditional methods discussed in this paper.

This paper is organized as follows. Section 2 presents the data set and the characteristics of both satellites (Sentinel-1 and TerraSAR-X). In Sections 3 and 4, we deeply explain the deep learning methods for SAR image translation, including the development of traditional methods and the creation of the proposed method. Section 5 describes the experiments based on an urban area using the traditional and proposed methods, while Section 6 discusses the advantages of the proposed method compared with the traditional methods. Finally, Section 7 concludes this paper and gives future research perspectives.

## 2. Data set

In the field of radar remote sensing, they are many satellites for different applications [16]. In this paper, we chose two typical satellite systems, Sentinel-1 and TerraSAR-X, which serve the same purpose but with different characteristics.

Sentinel-1 is a C-band SAR satellite system launched by ESA, whose missions include sea and land monitoring, emergency response after environmental disasters, and commercial applications [17]. In contrast, TerraSAR-X is an X-band Earth observation SAR satellite being operated under a public-private-partnership between the German Aerospace Center (DLR) and EADS Astrium (now Airbus), whose main features are its high resolution with excellent geometrical accuracy [18]. In our opinion, Sentinel-1 is a good option to generate large-scale SAR images, while TerraSAR-X is an adept solution for high resolution. To avoid being influenced by radar configurations, we try to keep the radar system parameters of two products as consistent as possible. A comparison of the radar parameters of two image products we used in this paper is shown in Table 1.

Table 1. Selected data set parameters

| SAR instrument | TerraSAR-X | Sentinel-1A |
| --- | --- | --- |
| Carrier frequency band | X-band | C-band |
| Product level | Level 1b | Level 1 |
| Instrument mode | High Resolution Spotlight | Interferometric Wide Swath |
| Polarization | VV | VV |
| Orbit branch | Descending | Ascending |
| Incidence angle | 39° | 30°-46° |
| Product type | Enhanced Ellipsoid Corrected (EEC) (amplitude data) | Ground Range Detected High Resolution (GRDH) (amplitude data) |
| Enhancement | Radiometrically enhanced | Multi-looked |
| Ground range resolution | 2.9 m | 20 m |
| Pixel spacing | 1.25 m | 10 m |
| Equivalent number of looks (range×azimuth) | 3.2×2.6 = 8.3 | 5×1 = 5 |
| Map projection | WGS-84 | WGS-84 |
| Acquisition date | 2013-04-29 | 2014-10-13 |
| Original full image size (cols×rows) | 9200×8000 | 34,255×18,893 |
| Used image sizes (cols×rows) | 6370×4320 | 1373×936 |

*2.1 Image quantization*

The amplitude of SAR image products is usually not in the range of [0, 255] which is the dynamic range where optical image products stay. The amplitude of SAR images relates with the radar cross section (RCS) and has a large dynamic range. There are many methods for SAR image quantization [19]. Because we need to use pre-trained neural networks designed for optical images, the SAR data should be scaled to the brightness range of optical pixels. In order to generate the SAR images with good visual effects, an 8-bit uniform quantization is applied in different brightness range. For Sentinel-1 images, the range is [0, 800] while for TerraSAR-X images it is [0, 570]. These parameters were defined by the brightness levels of our test data which contain 98% of the pixels in the pixel brightness histograms.

*2.2 Image co-registration*

The image translation between two different products should be done with co-registered image pairs. Fortunately, remote sensing products can be projected the same coordinates by using geo-coding. Geo-coding is a technique that yields every pixel its longitude and latitude on Earth. Thus, for each pixel, once its location is determined, the pixel information from both Sentinel-1 and TerraSAR-X images can be retrieved. In order that the two images have the same content and the same pixel size, the pixel spacing for both images is set to the same value, where the scale is 1:10. Finally, the interpolation and the co-registration are completed automatically in the QGIS software, which is an open source tool. In this software, the interpolation is based on IDW (Inverse Distance Weighted) method [20], and the co-registration relies on the annotation data of the image product resulting the accuracy of a few meters.

*2.3 Training data and test data*

The selection of a training data set and a test data set for quality control is a primary task in deep learning. There are several hyper-parameters to be determined and they can finally impact the capabilities of the trained networks. The selected patch size is one of the hyper-parameters that can affect both the final results and the amount of the training data. When the patch size is too large, the number of the training data becomes small, even the data augmentation can be applied. Based on the discoveries in [21], which yielded a best patch size for SAR image classification, we chose for our studies a patch size of 128×128 pixels [21]. Using an overlap of 50% between the tiled patches, we obtained 1860 patches for training and 224 patches for testing.

**3. Related work**

Deep learning has been widely used in the last years in computer vision, biology, medical imaging, and remote sensing. Although the theory of deep learning is not yet mature, its capabilities shown in numerous applications have attracted the attention of many researchers. Let us simply review the development of image translation with deep learning. In 2016, Gatys et al. demonstrated the power of Convolutional Neural Networks (CNNs) in creating fantastic artistic imagery. With a good understanding of the pre-trained VGG networks, they have achieved the style transfer and demonstrated that semantic exchange could be made by using neural networks. Since then, Neural Style Transfer has become a trending topic both in academic literature and industrial applications [22]. To accelerate the speed of Neural Style Transfer, a lot of follow-up studies were conducted. A typical one is Texture networks. With the appearance of GANs, several researchers turned to GANs to find more general methods without defining the texture. In this paper, we examine three typical methods, the method of Gatys *et al.* [9], Texture Networks [10] and Conditional GANs [13]. By analyzing their advantages and disadvantages in SAR image translations, we propose a new GAN-based framework which is the combination of the manifestations of SAR images in the VGG-19 network, the definition of texture content, and the WGAN method.

*3.1. VGG-19 network*

VGG-19 is a key tool to conduct style transfers. It is a pre-trained CNN model for large-scale visual recognition developed by Visual Geometry Group at the University of Oxford, which has achieved excellent performances in the ImageNet challenge. Gatys *et al.* [9] firstly introduced this CNN in their work. Then, the next studies were focused on the utilization of the outcomes of VGG-19. However, VGG-19 has been trained on the ImageNet dataset which is the collection of optical images. In order to find the capabilities of VGG-19 for SAR images, we first visualize the content of each layer in VGG-19 when the input is a SAR image and then analyze the meaning of each layer. The input SAR images are in the 8-bit dynamic range without histogram changes for fitting the optical type. There are 19 layers in the VGG-19 network, but the most commonly used layers are the layers after down-sampling, which are called ReLU1_1, ReLU2_1, ReLU3_1, ReLU4_1, and ReLU5_1. A visualization of SAR images via the VGG-19 layers is shown in Figure 1.

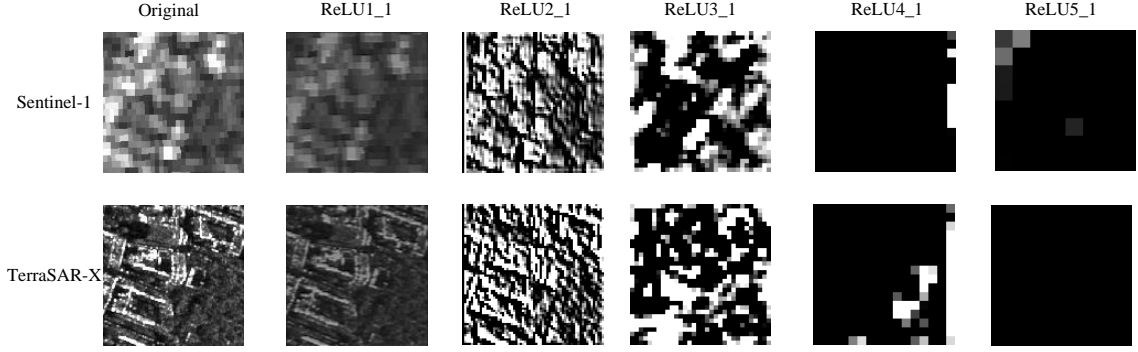

**Figure 1**. Visualization of Sentinel-1 and TerraSAR-X SAR images in the VGG-19 layers

As can be seen from Figure 1, the images in ReLU 1_1, ReLU 2_1, and ReLU 3_1 layers are quite different, while the images in ReLU 4_1 and ReLU5_1 of both two sensors are similar. According to the conception of deep learning, the higher layers contain higher semantic information [9], which supports the results in Figure 1. Therefore, Gatys *et al.* used the shallow (i.e., lower) layers as the components of texture and took the deep layers as the content information. However, we find that the ReLU5_1 images in both Sentinel-1 and TerraSAR-X are almost featureless. In another paper [23], the authors found that ReLU5_1 has real content for optical images. This may be because this training of VGG-19 is based on optical images. Whatever, we decide to ignore the ReLU5_1 layer in our algorithm in order to accelerate the computation. It will be discussed in the experiment part.

*3.2. Texture definion-Gram matrix*

The success of Gatys' paper is to some extent achieved by the introduction of a Gram matrix. If we regard the pixels of the feature map in each layer as a set of random variables, the Gram matrix is a kind of second-order moment. The Gram matrix in that paper is computed on the selected layers as described in Section 3.1. Assuming $L$ layers are selected and their corresponding number of feature maps is $N_l$, the Gram matrix of the $l^{th}$ layer is

$$\mathbf{G}^l = \frac{1}{M_l} \begin{bmatrix} \mathbf{F}^l_{1:}{}^T \\ \mathbf{F}^l_{2:}{}^T \\ \vdots \\ \mathbf{F}^l_{N_l:}{}^T \end{bmatrix} \begin{bmatrix} \mathbf{F}^l_{1:} & \mathbf{F}^l_{2:} & \cdots & \mathbf{F}^l_{N_l:} \end{bmatrix}, \tag{1}$$

where $\mathbf{F}^l_{i:}$ is the column vector generated from the $i^{th}$ feature map of layer $l$, and $M_l$ is the size number of each feature map in this layer. An element of the $N^l \times N^l$ Gram matrix is

$$\mathbf{G}^l_{ij} = \frac{1}{M_l} \sum_{k=1}^{M_l} \mathbf{F}^l_{ik} \mathbf{F}^l_{jk} = \frac{1}{M_l} \langle \mathbf{F}^l_{i:}, \mathbf{F}^l_{j:} \rangle, \tag{2}$$

where $\langle \cdot \rangle$ denotes the inner product. When we get the Gram matrices $\{\mathbf{G}^l\}_{l \in \mathbf{L}_{selected}}$, where $\mathbf{L}_{selected}$ is the set of the selected layers to define the texture information. Having the Gram matrices, the definition of the style difference between two images is

$$\mathcal{L}_{style} = \sum_{l \in \mathbf{L}_{selected}} w_l \|\hat{\mathbf{G}}^l - \mathbf{G}^l\|_F^2, \tag{3}$$

where $w_l$ is a kind of hyper-parameter define the weight of the style in the $l^{th}$ layer, $\hat{\mathbf{G}}^l$ is the Gram matrix of the being generated image in the $l^{th}$ layer, $\mathbf{G}^l$ is the corresponding term for the reference image, and $\|\cdot\|_F$ is the Frobenius norm of the matrices. In our case, the style image is no longer an artistic painting of art, and the Gram matrices did not perform well. Figure 2 shows the mismatch of utilizing these Gram matrices to translate between SAR images.

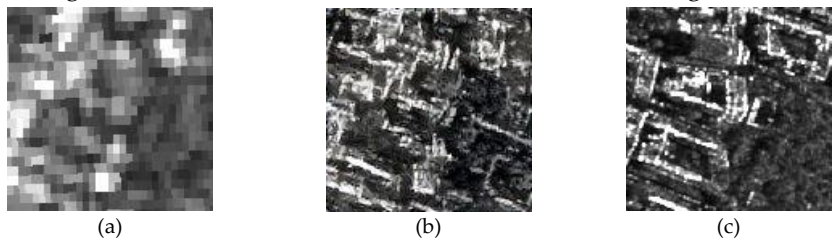

**Figure 2.** Experiment using the Gatys *et al.* method **(a)** content image (Sentinel-1) **(b)** transferred image (Gram matrix) **(c)** style image (TerraSAR-X)

Figure 2(b) contains many fake targets. For example, there is nothing at the lower right part of both Figure 2(a) and Figure 2(c), but some bright lines, usually from buildings, appear at that part of Figure 2(b). Besides, contrary to Figure 2(c), the layout of buildings in Figure 2(b) is hard to understand. In our experiment, the SAR data are depicting an urban area, where most targets are buildings. The city structure is quite different from the design of artistic works, which means the style definition should vary for different applications. Reflecting upon the Gram matrices, their format should be changed. The vectorization of the feature maps makes the Gram matrices fully blind to the arrangement information inside the maps [24]. To maintain the arrangement information, which is useful for urban area, we should discuss this arrangement information and how to make it suitable for our applications.

The arrangement most often indicates the placing of items according to a plan, but without necessarily modifying the items themselves. Thus, an image with arrangement information should contain similar items and the similar items are placed in different locations. When we tile the images into small pieces (called patches) according to the scheme they belong to, the small pieces should be similar. Their similarity can be determined by the Gram matrix, while the way to tile the image is the part of our approach. The manifestation of most objects of urban areas in remote sensing images is usually rectangular. Thus, the main outline of urban SAR images should be straight lines.

The Spatial Gram method is a good way to represent this arrangement texture, which defined by the self-similarity matrices themselves and by applying spatial transformations when generating these matrices. A Gram matrix is a measurement of the relationship of two matrices, and the spatial transformation determines which two. G. Berger *et al.* have proposed a series of CNN-based Spatial Gram matrices to define the texture information. Based on their ideas in [24], we apply a spatial transform tiling the feature map horizontally and vertically in different levels to represent the "straight" texture information.

As we have several options to tile an image, how to compute their Gram matrices to define the texture is still a question, either to add them or to regard them as parallel structures. When the Spatial Gram computation just has one element, it degenerates into the traditional Gram matrix like the one used by Gatys *et al.* But when it has too many elements, the ultimate configuration is that all the pixels are in the Gram matrix individually and it will lose its capability to generate diverse textures. A line, which is the basic unit of our images, can be determined by two parameters. Thus, we use the two orthogonal dimensions ($row$ and $col$), as two rows of the Spatial Gram matrix, and the spatial transform types as the columns. Thus, the Spatial Gram matrix we applied in this paper is

$$S_{spatial}^l = \begin{pmatrix} G_{row,2}^l & G_{row,4}^l & \cdots & G_{row,2^{7-l}}^l \\ G_{col,2}^l & G_{col,4}^l & \cdots & G_{col,2^{7-l}}^l \end{pmatrix}, \quad (4)$$

where the type of transformation is related to the size of the feature maps in this layer. $\Delta_l = \{2, \ldots, 2^{7-l}\}$ where the 7 is determined by the input size of patches (128×128), and $L_{seleted} = \{1, 2, 3\}$. $G_{row,\delta}^l$ and $G_{col,\delta}^l$ are two kinds of spatial transformation which is related to the dimensions $row$ and *col*, and the shifted amount $\delta$. Assuming the feature map is $\mathbf{F}^l$, and its transformations are $T(\mathbf{F}^l)$ where $T$ denotes the function of spatial transformation. For example, the spatial transformations of feature maps in the row dimension are defined as

$$T_{row,\delta}(\mathbf{F}^l) = \mathbf{F}^l(\delta:M, 1:N),$$
$$T_{row,-\delta}(\mathbf{F}^l) = \mathbf{F}^l(1:M-\delta, 1:N),$$
(5)

where $M, N$ are the height and width of the feature map $\mathbf{F}^l$. $T_{row,\delta}(\mathbf{F}^l)$ is the transformation on the row dimension. The vectorization of $T_{row,\delta}(\mathbf{F}^l)$ is written as $T_{row,\delta}(\mathbf{F}^l)_:$ which is the column vector. Having these definitions, $\mathbf{G}^l_{row,\delta}$ can be written as

$$\mathbf{G}^l_{row,\delta} = \frac{1}{M_l}\begin{bmatrix} T_{row,\delta}(\mathbf{F}^l_1)_:^T \\ T_{row,\delta}(\mathbf{F}^l_2)_:^T \\ \vdots \\ T_{row,\delta}(\mathbf{F}^l_{N_l})_:^T \end{bmatrix}\begin{bmatrix} T_{row,-\delta}(\mathbf{F}^l_1)_: & T_{row,-\delta}(\mathbf{F}^l_2)_: & \cdots & T_{row,-\delta}(\mathbf{F}^l_{N_l})_: \end{bmatrix},$$
(6)

where $\mathbf{G}^l_{row,\delta}$ can be written in the same way but the spatial transformation takes places in the row direction. Thus, the spatial style loss function is

$$\mathcal{L}_{style} = \sum_{l \in L_{selected}} w_l \|\hat{\mathbf{S}}^l_{spatial} - \mathbf{S}^l_{spatial}\|_F^2.$$
(7)

where the $\mathbf{S}^l_{spatial}$ if the spatial matrices of the target images and $\hat{\mathbf{S}}^l_{spatial}$ is for the generated image. The style loss function $\mathcal{L}_{style}$ is only dominated by the Spatial Gram matrices, it is not necessary to add the traditional Gram matrices because when $\delta$ is small, it is almost the same as the traditional one. Figure 3 shows the results applying the new Spatial Gram matrix.

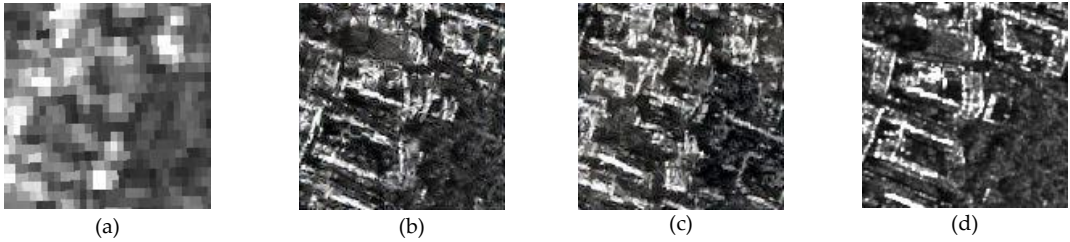

(a)　　　　　(b)　　　　　(c)　　　　　(d)

**Figure 3.** Experiment using Spatial Gram matrices (**a**) content image (Sentinel-1) (**b**) transferred image (Spatial Gram matrix) (**c**) transferred image (Gatys *et al.*'s Gram matrix) (**d**) style image (TerraSAR-X)

*3.3 Conditional generative adversarial networks*

The introduction of GANs is a milestone in deep learning, and it becomes popular where hundreds of papers were published under the name of GAN [25]. A conditional GAN makes a general GAN more useful because the inputs are no longer the noise but the things we can control. In our case, the conditional inputs are Sentinel-1 images. The conditional GANs have achieved impressive results on many image processing tasks, such as style transfer [26], supper-resolution [27], or other tasks [28], [29]. Isola *et al.* [13] summarized the tasks of image translation as *"pix2pixl"* translations and demonstrated the capabilities of conditional GANs in their paper. Inspired by their works, we modified the *"pix2pix"* framework by adding new discoveries about GANs and specific features of the SAR images translations. When we used the *"pix2pix"* framework in our application this failed. Figure 4 shows the overfitting of the *"pix2pix"* conditional GAN because the training set has good performances while the test set has bad results. Without any modification, we could not reach our goals. In the next section, we propose a new method to realize Sentinel-1 to TerraSAR-X image translations.

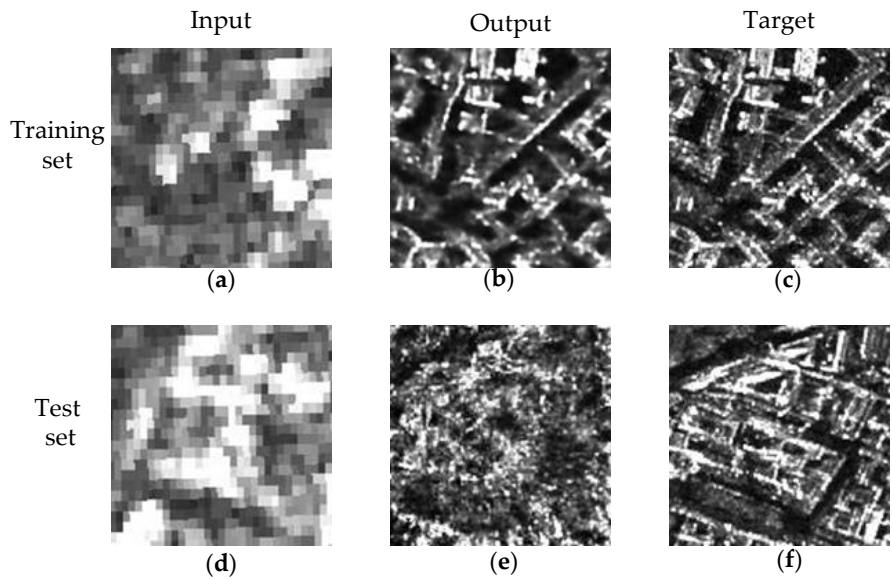

**Figure 4.** SAR image translation using the "*pix2pix*" framework in both training and test set (**a**) input image in the training set (**b**) GAN output of image (a) (**c**) target of image (a) (**d**) input image in the test set (**e**) GAN output of image (d) (**f**) target of image (d)

## 4. Method

Although the conditional GAN is overfitting in our case, it is still a good strategy to complete our task, which is to have a mapping function from Sentinel-1 to TerraSAR-X. In mathematical notation, it is

$$G: x \to y, \tag{8}$$

where $G$ is the mapping function, $x$ is a Sentinel-1 image, and $y$ is a TerraSAR-X image. Actually, this task can be achieved by designing a neural network and by presetting a loss function like traditional machine learning. Indeed, this idea has already been accomplished in [10] and [11]. However, the preset loss function is not general for all cases. A GAN provided an idea that the loss function is not preset, and it can be trained through a network which is called "Discriminator". The mapping function $G$ is realized through a "Generator" neural network.

In this paper, we use the concept of dialectics to unify the GANs and traditional neural networks. There is a triad in the system of dialectics, thesis, antithesis and synthesis, and they are regarded as a formula for the explanation of change. The formula is summarized as (1) a beginning proposition called a thesis, (2) a negation of that thesis called the antithesis, and (3) a synthesis whereby the two conflicting ideas are reconciled to form a new proposition [30]. We apply this formula to describe the change of image translation. The "Generator" network is regarded as thesis and it can be inherit the parameters from the previous thesis. In our case, the "Generator" inherits from the texture network. The "Discriminator" network acts as a negation of the "Generator". The synthesis is based on the law of the Negation of the Negation. Thus, we can generate a new "Generator" through the dialectical method. When the new data comes, it will enter the next state of changing and development. The global flowchart of our method is shown in Figure 5. There are two phrases, training phrase and operational phrase. The training phrase is the processing to generate a final generator, and the operational phrase applies the final generator to conduct the image translation task. In the following, we discuss the "Generator" network, the "Discriminator" network and the details to train them.

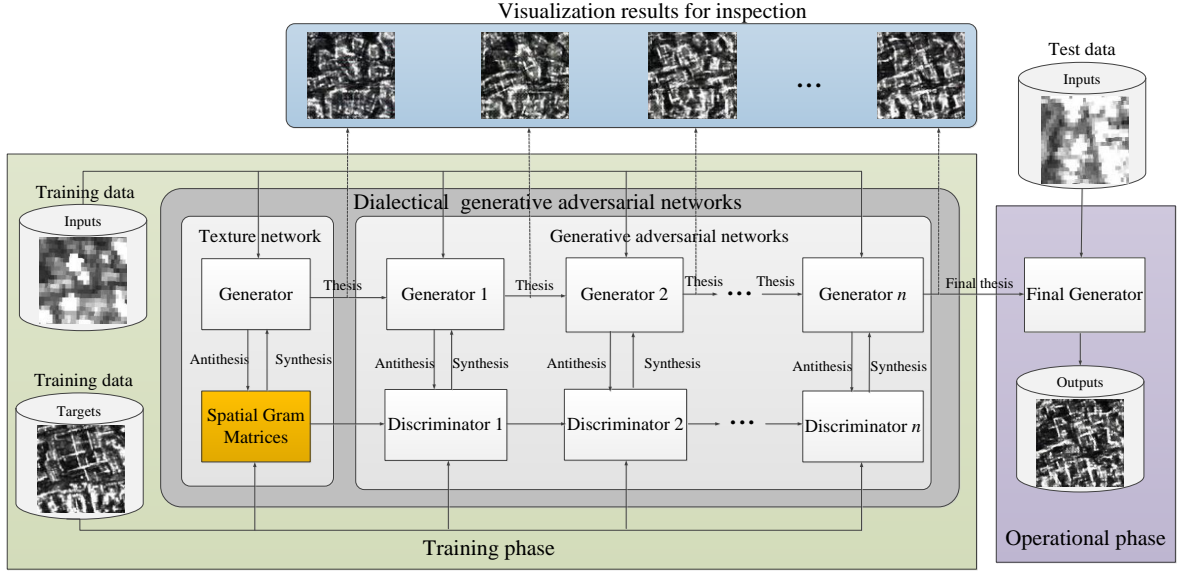

**Figure 5** Global flowchart of Dialectical GAN

*4.1 "Genertor" network – thesis*

The purpose of the generator is to generate an image $G(x)$ has the content of image $x$ and the style of image $y$. Thus, the loss function has two parts, content loss and texture loss, which is defined as

$$\mathcal{L}_{Generator} = \mathcal{L}_{content} + \lambda \mathcal{L}_{style} \\ = \sum_{l \in L_{content}} \left\| \mathbf{F}^l(G(x)) - \mathbf{F}^l(x) \right\|_F^2 + \lambda \sum_{l \in L_{style}} w_l \left\| \mathbf{S}_{spatial}^l(G(x)) - \mathbf{S}_{spatial}^l(y) \right\|_F^2, \quad (9)$$

where $\lambda$ is a regularization parameter, $\mathbf{F}^l(\cdot)$ are the feature maps of the $l^{th}$ layer of an image, $\mathbf{S}_{spatial}^l(\cdot)$ are the Spatial Gram matrices that were defined in Section 3.2. According to the discussion in Section 3.1, there is no information in "ReLU5_1". Therefore, we chose "ReLU4_1" as the content layer, and "ReLU1_1", "ReLU2_1" and "ReLU3_1" as the style layers. Consequently, $\mathbf{L}_{content} = \{4\}$, and $\mathbf{L}_{style} = \{1, 2, 3\}$.

$G$ can be any kind of functions, it can be as simple as a linear function or as complex as a multiple composition of non-linear functions. As a powerful tool to approximate functions [31] [32], deep neural networks are used as our notation of $G$ in this paper. The input and the target images, $x$ and $y$, are from different SAR sensors, but they are observing the same test site. The properties of SAR systems result in their own characteristics of image representation, such as final resolution, polarization response, and the dynamic ranges. But the same observed area makes them share identical compounds. Regardless of the changes in time, $x$ and $y$ are generated from identical objects. For the analysis of our input and target images, there are plenty of network structures that solve this problem.

Previous related works [28] [33] have used an encoder-decoder network [34] where the input image is compressed in down-sampled layers and then be expanded again in up-sampled layers where the process is reversed. The main problem of this structure is whether the information is preserved in the down-sampled layers. Based on the discussion in [13], we chose the "U-Net" network to execute our tasks. The "U-Net" is very well known for its skip connections which are a way to protect the information from loss during transport in neural networks. According to the behavior of our SAR images in the VGG-19 network, we set the "U-Net" to 6 layers. The structure of the network we used is shown in Figure 5.

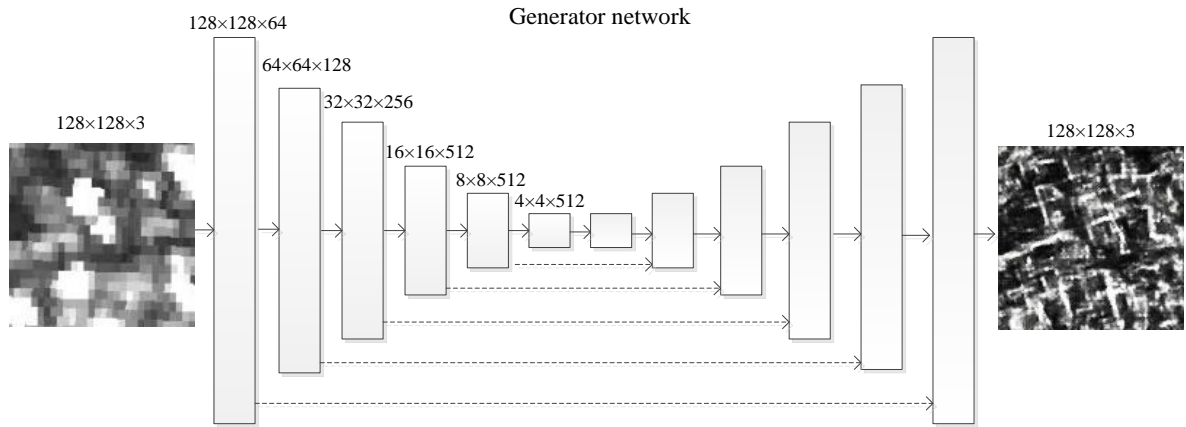

**Figure 6.** Architecture of the "U-Net" Generator network

Although the network in Figure 5 has too many elements and is hard to be trained, we think it is necessary to use a deep network because the architecture of a network can affect its expressiveness of complex functions. Maybe there will be more efficient methods to approximate the mapping function, but this is not the topic of this paper. Our goal is a powerful tool to describe the mapping from Sentinel-1 to TerraSAR-X where the solution is a deep neural network.

*4.2 "Discriminator" network – antithesis*

A deep neural network is a suitable solution, but on the other hand, it can also easily generate non-target results. Based on the concept of dialectics, when the appearance is not fit for the conception, it is needed to deny the existence of this thing. In this case, it is the negation of the generated images. In other words, we need a loss function yielding a small value when the output equals the target while yielding a high value when the two are different. Usually, the loss function is predefined. For example, the most common loss function, Mean Squared Error (MSE), is a preinstalled function which is defined as

$$MSE = \frac{1}{N}\sum_{i=1}^{N}\left(\mathbf{Y}^i - \hat{\mathbf{Y}}^i\right)^2, \tag{10}$$

where $\hat{\mathbf{Y}}$ is the generated vector of $\mathbf{Y}$ whose elements are $\mathbf{Y}^i$. When computing the MSE function, it outputs a scalar value to describe the similarity of the input and the target. But it is predefined, and the only freedom are the input data. How it relates to the negation of the generated images is still a question. There are three steps to solve the problem. First, the loss function should criticize the existence of $\hat{\mathbf{Y}}$, so it has a term $-\hat{\mathbf{Y}}$. Second, it should approve the subsistence of $\mathbf{Y}$, the target; thus, the term $\mathbf{Y}$ shall appear. Third, the square operator makes sure the function is a kind of distance. Through these three steps, the MSE has accomplished the negation of the generated vectors or images. When the generated image differs from the target image, the distance is large. When the generated image is the target image itself, their distance shall be zero. In contrast, a large distance shall be generated when the input is markedly different from the target to lead to better negation.

It is reasonable to expect that the loss function is a kind of distance function because the distance space is a weak assumption for the space of generated images. For instance, the loss function in (9) is another kind of distance compared with the MSE that directly computes pixel values. However, it is hard to find a unique common distance because our tasks differ while the distance remains invariant. Using a neural network scheme to train a distance is a good choice. Fortunately, the appearance of GANs has supported us solutions to find the proper distances. In GAN systems, the negation of generated images is processed in the loss function of the "Discriminator". The discriminator is a mapping function, or a neural network to describe the existence of the input image. However, the properties of the discriminator have been little discussed. In this paper, we try to use the theory of metric spaces to discuss this question.

Assuming that the distance in the image domain $M_1$ is $d_1(\cdot)$ and the distance in the discriminator domain $M_2$ is $d_2(\cdot)$, the discriminator is the map $D: M_1 \to M_1$ [35]. The distance of the conditional case, which is also the contradiction between two images, can be defined as

$$\mathcal{L}_{\text{contradiction}} = d_2\big(D(y|x), D(G(x)|x)\big), \tag{11}$$

where $D(\cdot|x)$ is the discriminator of an image under the condition that the input is $x$. If $D(\cdot)$ is a map to map the image to itself, and $d_2(\cdot)$ is the Frobenius norm, the contradiction becomes

$$\mathcal{L}_{\text{contradiction}} = \|y - G(x)\|_F, \tag{12}$$

which is the $L_1$ norm that usually acts as a loss function in machine learning. This is one case of a determined map. As for a training map function, the most important thing is to design its format. If we still set $d_2(\cdot)$ as the Frobenius norm, the distance of the discriminator becomes

$$\mathcal{L}_{\text{contradiction}} = \|D(y|x) - D(G(x)|x)\|_F, \tag{13}$$

when the discriminator is a predefined network such as the Spatial Gram matrix, we conclude that the loss function in (9) can be regarded as a specific case of (13).

If the range of $d_2(\cdot)$ is [0,1], it is considered that the output is the possibility of being real. There are many concepts to re-unite the formats of different loss functions. In $f$-GAN [36], the loss functions are regarded as $f$-divergences, which are the measurements for the similarity of two distributions. However, the drawback of divergences is that they don't satisfy the triangle inequality and the symmetry which are requirements of distance functions [37]. In LSGAN [38], the least squares method is used to measure the output of the discriminator. In this method, the generated images are in an inner product space which is also a metric space. Therefore, we infer that the contradiction of the real image and the generated image should be contained in a function that can define the distance of some metric space, and the map $D$ should be constrained. One constraint of $D$ is that the range of $D$ should be bounded because we compute it in a computer. Or it will become an infinite number. Second, $D$ should be continuous, even uniformly continuous, because the gradient descent algorithms may fail when the loss function is not continuous. In WGAN, the Wasserstein distance is used, where the Lipschitz-continuous map ensures the property of uniformly continuous. In this paper, we focus on the WGAN framework.

When $d_2(\cdot)$ is the Wasserstein distance [39], the loss function of the discriminator becomes

$$\mathcal{L}_{\text{discriminator}} = W\big(D(y|x), D(G(x)|x)\big), \tag{14}$$

where $W(\cdot)$ is the Wasserstein distance function which behaves better than the $f - divergence$ being used in traditional GANs. The realization of the Wasserstein distance enforces a Lipschitz constraint on the Discriminator. In the WGAN-GP framework [40], the Lipschitz constraint is realized by enforcing a soft version of the constraint with a penalty on the gradient norm for random samples $\hat{x} \sim \mathbb{P}_{\hat{x}}$, where $\hat{x} = \epsilon y + (1 - \epsilon)G(x)$. Based on the conclusions in WGAN [40], the maximum of the Wasserstein distance between $\mathbb{P}_{r,y|x}$ and $\mathbb{P}_{g,x}$ becomes

$$D^* = \max_D(\mathcal{L}_{\text{discriminator}}) = \min_D \begin{pmatrix} \mathbb{E}_{G(x) \sim \mathbb{P}_{g,x}, x \sim \mathbb{P}_{r,x}}[D(G(x)|x)] - \mathbb{E}_{y \sim \mathbb{P}_{r,y}, x \sim \mathbb{P}_{r,x}}[D(y|x)] \\ + \lambda_{gp} \mathbb{E}_{\hat{x} \sim \mathbb{P}_{\hat{x}}}[(\|\nabla_{\hat{x}} D(\hat{x}|x)\|_2 - 1)^2] \end{pmatrix}, \tag{15}$$

where $D^*$ is the best discriminator, $\mathbb{P}_{r,y|x}$ is the distribution of given real images, $\mathbb{P}_{g,x}$ is the distribution of generated images and $\nabla_{\hat{x}} D(\hat{x}|x)$ is the gradient of the discriminator $D(\cdot|\cdot)$. When adding the penalty of the distance between the normal of $\nabla_{\hat{x}} D(\hat{x}|x)$ and 1 in the loss function, the Discriminator is forced to become a $1 - Lipschitz$ function. $\lambda_{gp}$ is usually set to 10 according to the experiments conducted in [40]. Intuitively, the removal of the absolute operator ensures the continuity of the derivation of the loss function at the origin. The $1 - Lipschitz$ constraint limits the normal of the derivation from growing too large, which is a way to increase the distance but not in the way we want.

Once the loss function is determined, the next step is to design the architecture of $D(\cdot|x)$ that can be easily trained for computers. Considering the ready-made function already discussed in the previous section, the loss function of style defined by Gram matrices is a good choice because it can be regarded as processing on a Markov random field [13] [26]. The *"pix2pix"* summarized it as the *"PatchGAN"* whose input is the combination of $x$ and $y$. The architecture of the discriminator is shown in Figure 7.

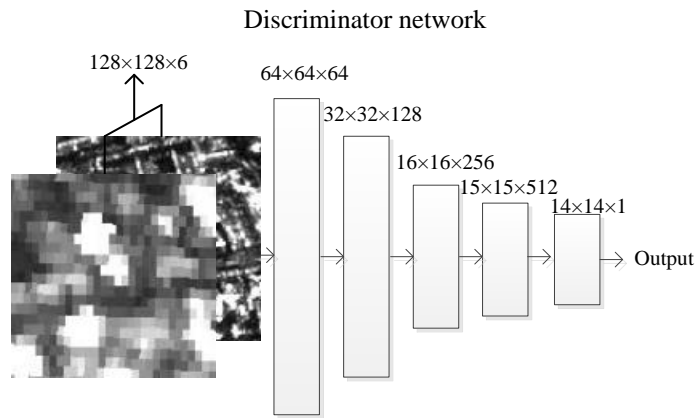

**Figure 7.** Architecture of PatchGAN Discriminator network

*4.3 Dialectical Generative adversarial network– synthesis*

According to the dialectic, the third step is the negation of the negation. The negation of the generated image is described by the loss function of the discriminator. Thus, the negation of negation should be the negation of the loss function of the discriminator. The negation is trying to make the distance defined by the discriminator to become larger, while the negation of negation should make the distance smaller. In our WGAN framework, the negation is defined by equation (15). Thus, the negation of negation can be realized by maximizing it. Therefore, the maximization of the loss function in (15) is the negation of negation. At the last step of the dialectic, the negation of negation should be combined with the thesis to form a synthesis.

The thesis can be regarded as a synthesis from the former dialectics. For example, the *"pix2pix"* used the $L_1$ norm as their thesis, and the SRGAN used the Gram matrices on layer 5 of the VGG-19 network as their thesis. These initial loss functions are distance functions and contain the negation of the generated images. In this paper, we start from the thesis defined by a Spatial Gram matrix. In other words, we set the initial loss function as defined in (9). The negation of negation is the maximization of (15). Therefore, the synthesis of our "Dialectical GAN" is the combination of (9) and (15). Reducing the terms in (15) that independent of "Generator" networks, the loss function of the "Dialectical GAN" becomes

$$\begin{aligned}\mathcal{L}_{Generator}^{GAN} &= \mathcal{L}_{Generator} - \lambda_{GAN}\mathcal{L}_{critical}\\ &= \mathcal{L}_{content} + \lambda\mathcal{L}_{style} - \lambda_{GAN}\mathop{\mathbb{E}}_{G(x)\sim\mathbb{P}_{g,x}, x\sim\mathbb{P}_{r,x}}[D(G(x)|x)]\end{aligned} \quad (16)$$

To optimize this new loss function, we need four steps: set up the generator, update the discriminator, update the generator and iterate.
- *Step 1*, having a Generator $G(\cdot)$ and an input image $x$, use the to generate $G(x)$, and then run the Discriminator $D(\cdot|\cdot)$.
- *Step 2*, use gradient descent methods to update $D(\cdot|\cdot)$ following (16)
- *Step 3*, use gradient descent methods to update $G(\cdot)$ following (15).
- *Step 4*, repeat *Step 1* and *Step 3* until the stopping condition is met.

Then the training of the Dialectical GAN is completed. Every loop can be considered as a realization of the dialectics. The basic framework is based on the WGAN-GP. As for the mathematical analysis of the GANs and deep learnings, please refer to [41], [42], [43]. Although the Deep Learning still looks like a "black box", we tried to provide a logical analysis of it and attempted to achieve "real" artificial intelligence with the capabilities of dialectics.

## 5. Experiments

The data used for demonstration has already been described in Section 2. Based on the method proposed in Section 4, the GAN network used in this paper has two neural networks, Generator and Discriminator. The Generator is a "U-Net" with 6 layers, and the Discriminator is a "PatchGAN" convolutional neural network with 4 layers. In total, we had 1860 image pair-patches in the training data set and 224 image pair-patches in the test data set. With these data sets, the training took two days on a laptop with Intel Xeon CPU E3, an NVidia Q2000M GPU and 64 GB of memory. We conducted three experiments with respect to the following networks further presented below.

### 5.1. SAR images in VGG-19 networks

VGG-19 has an essential role in this paper because its layers are the components of the texture information determined by a Gram matrix. Besides, the selection of the content layer is a new problem for SAR images. First, we compared the differences between Sentinel-1 and TerraSAR-X images in each layer. Two image patch-pairs are the inputs in the VGG-19 networks and their intermediate results are shown in Figure 8.

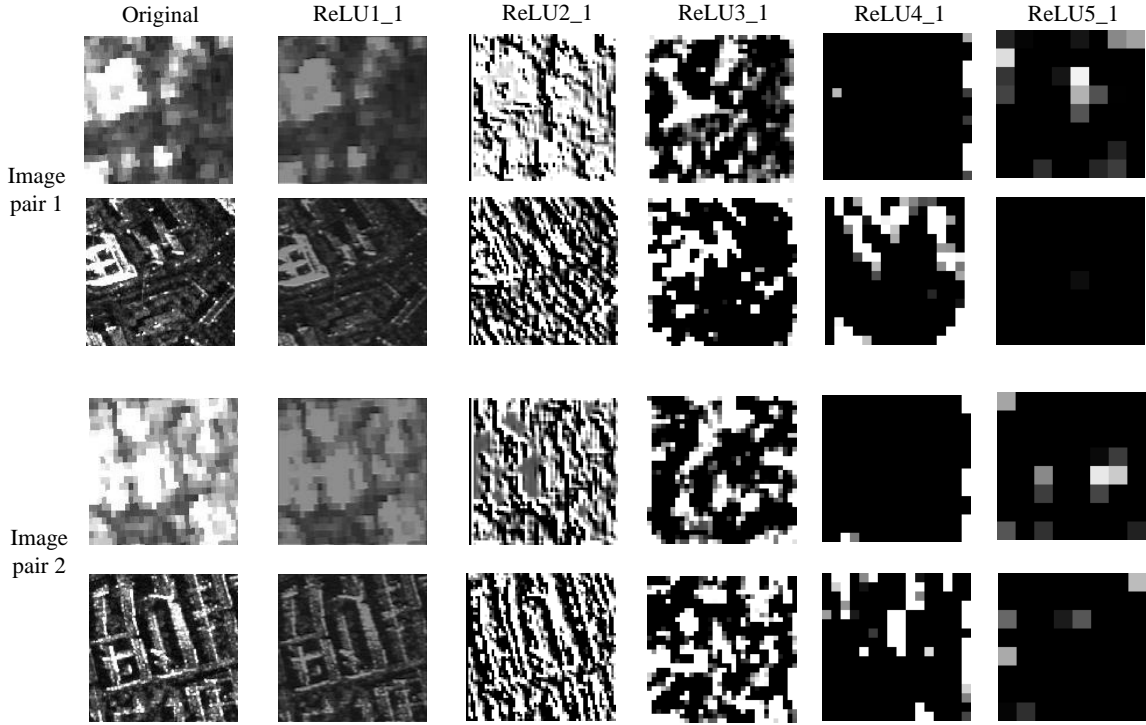

**Figure 8.** Two image patch-pairs input to in the VGG-19 networks and their intermediate results

Visually, the images of the ReLU4_1 layer have common parts. But this is not sufficient, and we decided to introduce the MSE and the Structural Similarity Index (SSIM) [44] in order to compare the image in different layers. The MSE is defined as:

$$MSE^l = \frac{1}{(M^l)^2 N^l} \sum_{k=0}^{N^l-1} \sum_{i=0}^{M^l-1} \sum_{j=0}^{M^l-1} \left[x_k^l(i,j) - y_k^l(i,j)\right]^2, \qquad (17)$$

where $M^l$ is the size of the feature maps in $l^{th}$ layer, $N^l$ is the number of the feature maps in $l^{th}$ layer, $x_k^l(i,j)$ is the pixel value of $(i,j)$ in the $k^{th}$ feature map of the $l^{th}$ layer of a Sentinel-1 image, and $y_k^l(i,j)$ is the counterpart of a TerraSAR-X image. In order to overcome the drawbacks of the MSE, we applied the SSIM whose definition is

$$SSIM(x,y) = \frac{(2\mu_x\mu_y + c_1)(2\sigma_{xy} + c_2)}{(\mu_x^2 + \mu_y^2 + c_1)(\sigma_x^2 + \sigma_y^2 + c_2)}, \tag{18}$$

where $\mu_x$ and $\sigma_x$ are the mean value and the standard deviation of image $x$; the same to applies to $y$. $c_1$ and $c_2$ are two constants related with the dynamic range of the pixel values. For more details, we refer the reader to [44]. The SSIM values range between -1 and 1, where 1 indicates perfect similarity. The evaluation results with the two indicators are shown in Table 2.

Table 2. Evaluation results with MSE and SSIM

| Layers | MSE | SSIM |
| --- | --- | --- |
| ReLU1_1 | 0.1616 | 0.4269 |
| ReLU2_1 | 0.5553 | 0.0566 |
| ReLU3_1 | 0.5786 | 0.2115 |
| ReLU4_1 | 0.3803 | 0.7515 |
| ReLU5_1 | 0.2273 | 0.7637 |

Although ReLU5_1 has the best performance with two indicators, we still ignore this layer due to the poor diversity in this layer. Excluding ReLU5_1, the ReLU4_1 layer gives us the best result. Therefore, the ReLU4_1 is chosen as the content layer, and the first three layers are used to define texture information.

*5.2. Gram martrices vs. Sptatial Gram martrices*

A Spatial Gram matrix is an extension of a Gram matrix, which is used to describe the texture information and is good at representing arrangement information. In Section 3.2, we have shown the visual difference between two style definitions. In this experiment, we used the quantity indicators to evaluate the two methods. Two image patch-pairs were chosen to conduct the comparison, whose results are shown in Figure 9. In order to evaluate the image quality of the SAR images, we introduce the equivalent numbers of looks (ENL), which act as a contrast factor to represent the image resolutions approximately. A higher ENL value indicates that the image is smooth while a lower value means that the image is in high resolution [45]. For our case, we need high-resolution images and as a result, the lower their ENL value, the better. The definition of ENL is

$$ENL = \frac{\mu^2}{\sigma^2}, \tag{19}$$

where $\mu$ is the mean value of the image patch, and $\sigma$ is its standard deviation.

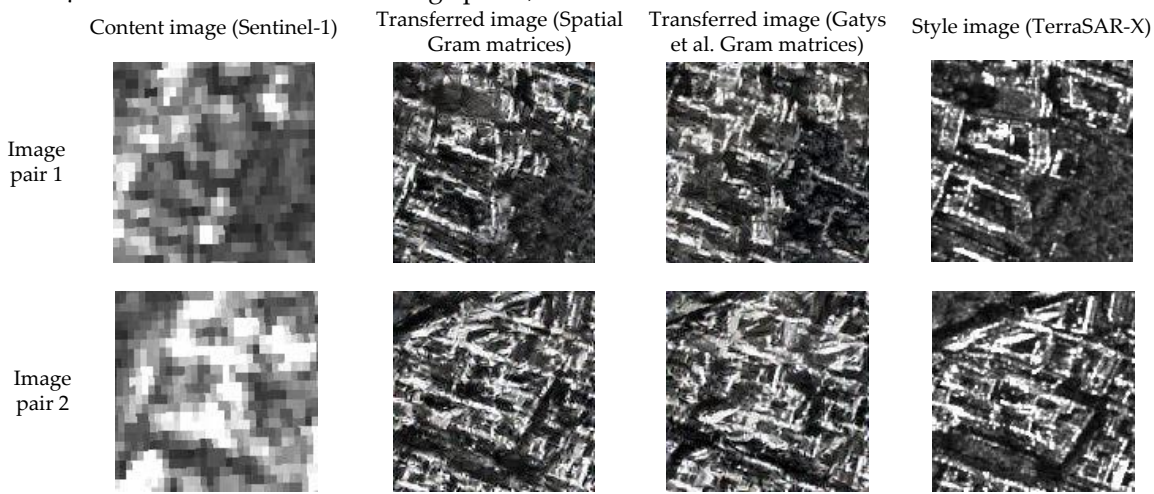

Figure 9. Comparison between a Spatial Gram matrix and a Gatys *et al.* Gram matrix in two patch-pairs

Table 3. Evaluation of two methods for both image pairs 1 and 2

| Image pairs | Methods | MSE | SSIM | ENL |
|---|---|---|---|---|
| 1 | Gatys *et al.* Gram | 0.3182 | 0.0925 | **1.8286** |
|   | Spatial Gram | **0.2762** | **0.1888** | 2.0951 |
| 2 | Gatys *et al.* Gram | 0.3795 | 0.0569 | 2.0389 |
|   | Spatial Gram | **0.3642** | **0.0700** | 1.9055 |

As can be seen from Figure 9 and Table 3, the Spatial Gram method performs better than Gatys *et al.*'s method, both visually and according to evaluation indicators. However, the ENL of image pair 1 indicates that Gatys *et al.*'s method is better. To solve this problem, we need more experiments. Because the traditional generative model regards every pixel as a random pixel and ignores the relationships among neighboring pixels, its computing efficiency is limited. Nevertheless, a Spatial Gram matrix is a good tool to determine the image style for our cases. In the next subsection, we abandon the Gatys *et al.*'s method and replaced it with a "U-net" network to generate the enhanced images. This method is called "Texture network".

*5.3. Spatial Gram matrices vs. traditional GANs*

The texture network moves the computational burden to a learning stage and no longer needs the style images as an aide to produce an image because the style information is already mapped in the network through the learning steps. Although the feed-forward network supersedes the solution of random matrices, the loss function is still the same. According to the above experiments, the Spatial Gram matrix is the winner of the determinate loss function.

In contrast to the determinate one, other researchers found that the loss function can also be learned, though the Spatial Gram matrix is also learned from the VGG-19 network. Nonetheless, the learning of the loss function enables the definition of image style to become more optional. We use the WGAN-GP framework to represent this kind of idea, which is the most stable one among the GAN family. The results of the texture network and the WGAN-GP are compared in Figure 10 and the evaluation results are listed in Table 4. The test set components in Table 4 are the average performances of images in whole test set.

The texture network and the WGAN-GP are fast ways to conduct style transfer. According to the values in Table 4, we conclude that the WGAN-GP has a better performance than the texture network method with the given indicators. However, the WGAN-GP is not able to preserve the content information of Sentinel-1 and its output images are muddled without obvious structures like the texture network. Although texture network has no good performance in the evaluation system, it has preferable visual effect in contrast to the WGAN-GP. How to balance the indicator values and the visual performance is a crucial problem. The texture information is defined by the VGG-19 network which has been trained by optical images. Thus, we have grounds to believe that there is texture information that cannot be described by Spatial Gram matrices. In a following experiment, we will compare the texture network with the proposed Dialectical GAN.

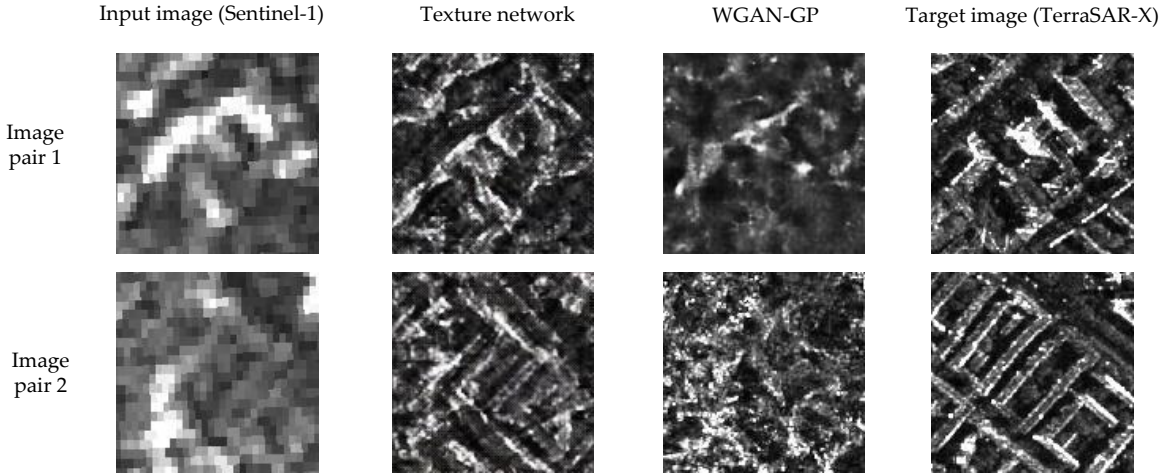

**Figure 10**. Comparison between Texture network and WGAN-GP for two patch-pairs

**Table 4.** Evaluation of Texture network and WGAN-GP in both image pair 1 and 2

| Image pairs | Methods | MSE | SSIM | ENL |
|---|---|---|---|---|
| 1 | Texture network | 0.3265 | 0.0614 | **1.3932** |
|   | WGAN-GP | **0.2464** | **0.1993** | 2.8725 |
| 2 | Texture network | 0.3396 | 0.0766 | **1.6269** |
|   | WGAN-GP | **0.2515** | **0.2058** | 3.5205 |
| Test set | Texture network | 0.3544 | 0.0596 | **1.7005** |
|   | WGAN-GP | **0.2632** | **0.2117** | 3.3299 |

*5.4. Dialectical GAN vs. Spatial Gram matrices*

The texture network defined the texture information in a determinate way while the WGAN-GP uses a flexible method to describe the difference between generative images and target images. In this paper, we proposed a new method that combines a determinate way and a flexible way to enhance the generative images, and we called it "Dialectical-GAN" because the idea is enlightened by the dialectical logic. The Dialectical-AN initializes its loss function with the Spatial Gram matrix that was found a good way to describe the texture information of urban area and the content loss defined by the ReLU4_1 layer of the VGG-19 network. Through the training of the Dialectical GAN, new texture information can be learned and represented in the "Discriminator" network. The comparison between a "Dialectical-GAN" and the texture network with a Spatial Gram loss function are shown in Figure 11 and Table 5.

Both visual performance (Figure 11) and the indicator analysis (Table 5) proved that our method is better than the texture network. However, these experiments all remained limited to the patch level, and the figures of a whole scene have not yet been considered. Therefore, we show the entire image composited with every path to check the overall performance and to estimate the relationship between neighboring patches.

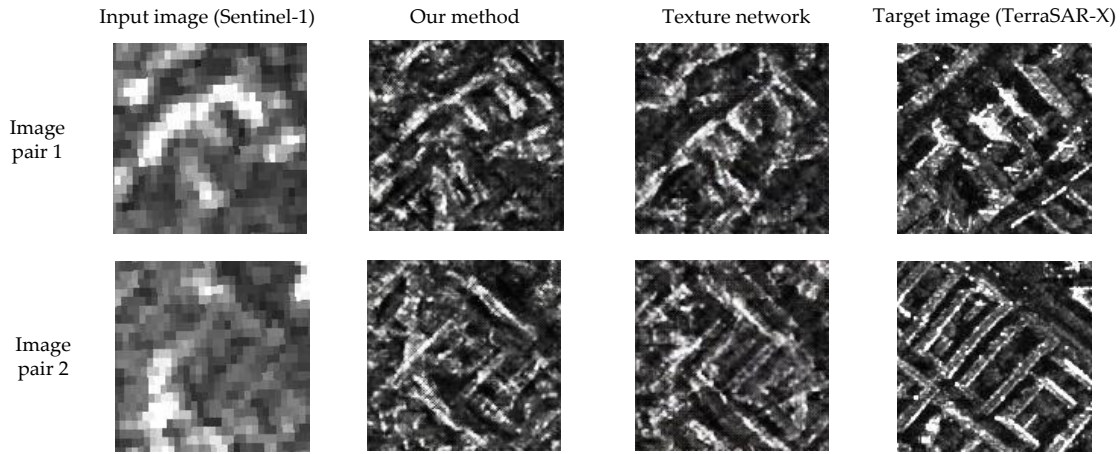

**Figure 11**. Comparison between Dialectical GAN and Texture network for two image patch-pairs

**Table 5.** Evaluation of Texture network and Dialectical GAN for both image pairs 1 and 2

| Image pairs | Methods | MSE | SSIM | ENL |
|---|---|---|---|---|
| 1 | Texture network | 0.3264 | 0.0614 | **1.3933** |
|   | Dialectical GAN | **0.3291** | **0.0884** | 1.5885 |
| 2 | Texture network | 0.3396 | **0.0766** | **1.6270** |
|   | Dialectical GAN | **0.3310** | 0.0505 | 1.8147 |
| Test set | Texture network | 0.3544 | 0.0596 | **1.7005** |
|   | Dialectical GAN | **0.3383** | **0.0769** | 1.8804 |

## 5.5. Overall visual perfomance

One of the most important merits of remote sensing images are their large-scale observations. In this section, we are discussing how a remote sensing image looks when its patches are processed by the selected neural networks. A full image is generated by concatenating the small processed patches to produce a final image. In this paper, we focus on three networks, the texture network with a Spatial Gram matrix, the WGAN-GP method, and our "Dialectical GAN" method. They are shown in Figure 12, Figure 13, and Figure 14, respectively. As for the overall visual performance, we consider that the Dialectical GAN has the best subjective visual performances.

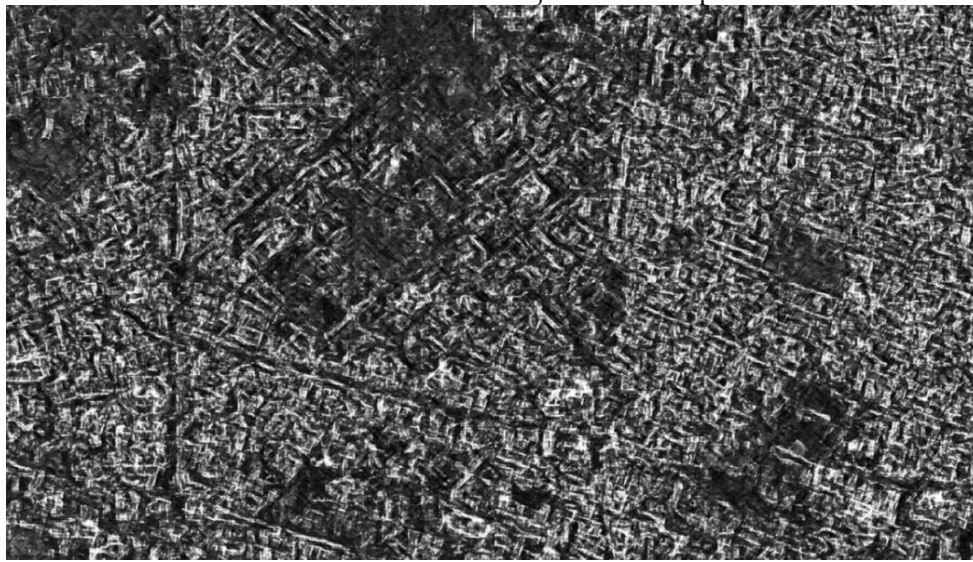

**Figure 12.** The overall results of a Dialectical-GAN

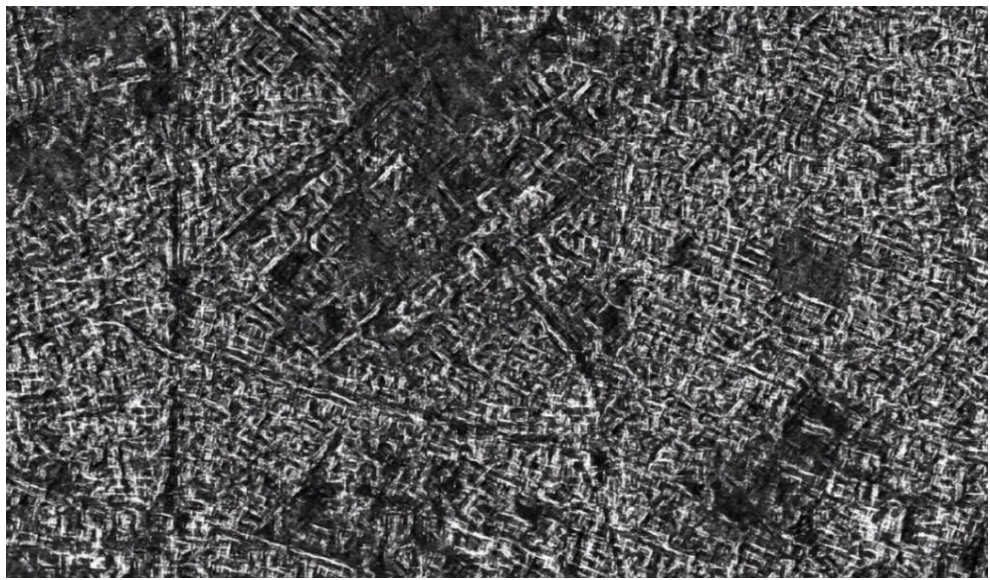

**Figure 13.** The overall results of a texture network

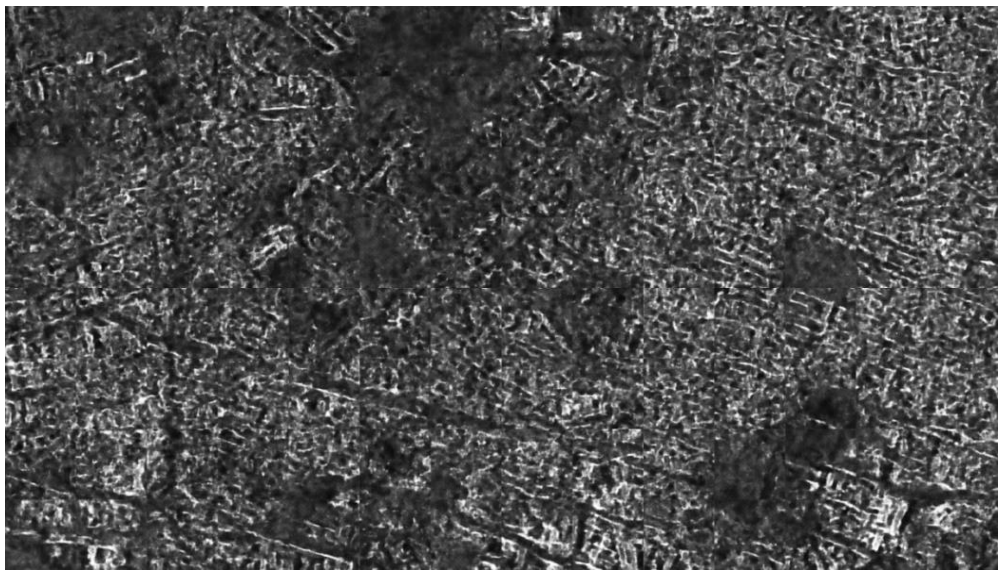

**Figure 14.** The overall results of a WGAN-GP (L1 +WGAN-GP)

The SAR image translation results compared with inputs and outputs image are shown in Figure 15. First, we can see the entire effect of the image translation.in the Munich urban area. To display detail results, we have three bounding box with different colors (Red, Green and Yellow) to extract the patches from the full image. They are in Figure 15(d).

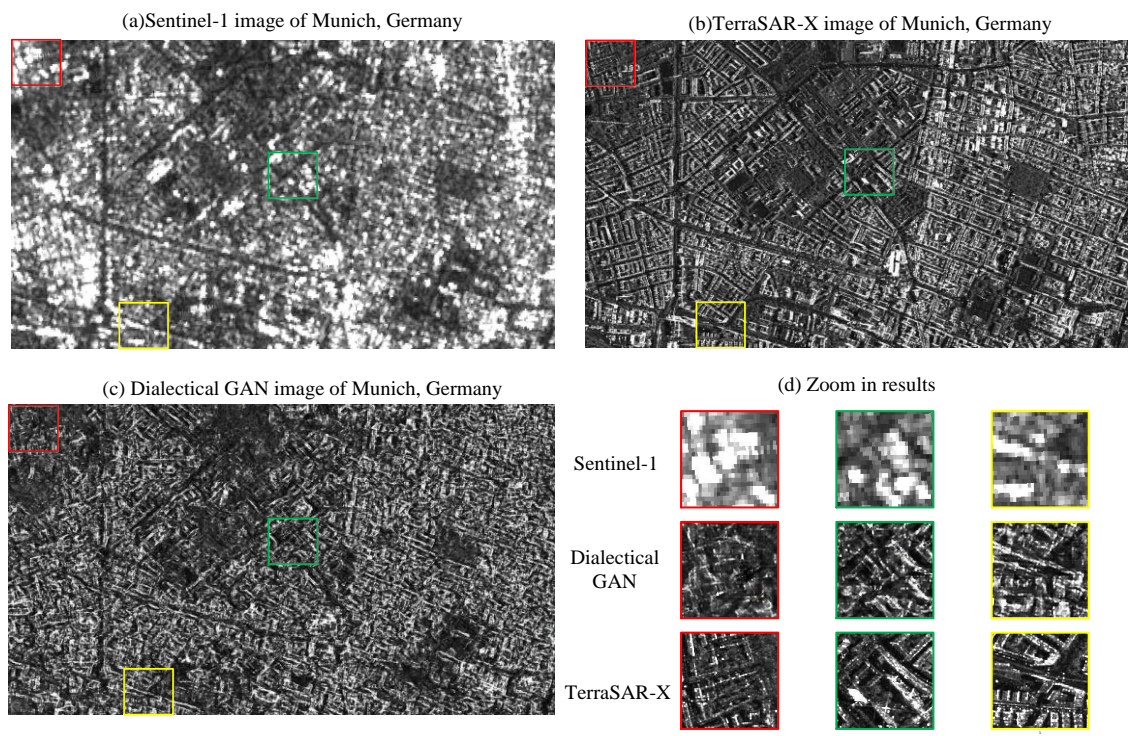

**Figure 15** Overall visual performance of Dialectical GAN compared with Sentinel-1 and TerraSAR-X images (a) Sentinel-1 image (b) TerraSAR-X image (c) Dialectical GAN image (d) Zoom in results

## 6. Discussion

Compared with traditional image enhancement methods, deep learning is an end-to-end method that is quite easy to be implemented. Deep learning has excellent performances and is standing out among the machine learning algorithms, especially in the case of big data. Solutions for remote sensing applications were discovered by the advent of deep learning. More importantly, deep learning is now playing a crucial part in transferring the style of images.

Concerning SAR image translation, little attention has been focused on it and the performances of deep learning on this topic are still unknown. The task that this paper addresses is related with super-resolution tasks, but our image pairs are not of the same appearances due to the differences in incidence angles, radar polarization, and acquisition times. From this aspect, our task belongs to style transfer to some extent, like generating a piece of artistic painting without the constraint that two images should be focused on same objects. Therefore, the SAR image translation is a mix of super-resolution and style transfer and has never been focused in the conception of deep learning.

From Gatys *et al.*'s method to GAN frameworks, we have tested the capabilities of deep learning in translating Sentinel-1 images to TerraSAR-X images. The resulting images of Gatys *et al.*'s method are of high quality but they don't preserve well the structure information, which is an essential characteristic of remote sensing SAR images, especially for urban areas. The improvement can be accomplished by introducing Spatial Gram matrices instead of the traditional ones in the loss function. A Spatial Gram matrix is a compromise between the arrangement structure and the freedom of style. In this paper, we compose Gram matrices computed in spatial shifting mode as a new matrix-vector for each layer. The spatial matrix is a good indicator to describe arrangement structures such as buildings and roads. However, our loss function modifications can only solve the style presentation problem, but the high computation effort still limits the applications of image translation for remote sensing. Fortunately, deep neural networks are a powerful tool for fitting complicated functions that provides solutions to speed up the image translation. Instead of taking every pixel as a random variable, a deep neural network regards an image as an input of the system, and the only thing the deep learning can do is to approximate the mapping function. That is to say, the deep neural network is a generator, and the Spatial Gram matrix is used to define the loss function.

The GAN framework gives us a new concept of a loss function which can also be defined by a neural network called discriminator. We assume that the GAN framework has a dialectical logical structure and explained it in a triad. However, due to the arbitrariness of a neural work and the limitation of the training data, a GAN is hard to train and cannot achieve good performances for our applications. Considering the diversity of GANs and the determinacy of Spatial Gram matrices, we proposed a new method that combines their advantages together. With the initial loss function defined by Spatial Gram matrices, our GAN system updates its discriminator and generator to make the output image as "true" as possible. The Spatial Gram loss function works well, but we still believe that there are other functions to determine the style of a given image. Using a combination framework, our system is able to generate high-quality SAR images and to improve the resolutions of Sentinel-1 images without the need for large amounts of data.

To appraise the generated images, we used three indicators, MSE, SSIM and ENL. The comparison experiments show that the Spatial Gram matrix is better than the traditional Gram matrix. A WGAN-GP without any initial loss function didn't perform well in contrast to the Spatial Gram matrix method. With the support of Spatial Gram matrices, the new WGAN-GP that we proposed is the best of these three methods, both in visual performance and by quantitative measurements (using the three indicators). Besides, we have tested the overall visual performance rather than to stay on image patch level. It is a new attempt for deep learning to perform the image transfer task in this way. The same results occurred when full images are considered and the new proposed method outperforms the existing ones.

## 7. Conclusions

In this paper, a "Dialectical GAN" based on Spatial Gram matrices and a WGAN-GP framework is proposed to conduct the SAR image transfer task from Sentinel-1 to TerraSAR-X images. By analyzing the behavior of SAR image in the VGG-19 pre-trained network, we have found that the relationship between two source images is maintained in the higher layers of the VGG-19 network, which is the foundation of changing the "style" of images. In remote sensing usually the urban areas are dominated by buildings and roads and, based on this observation, the Spatial Gram matrixes are a very good metric to describe the "style" information of these areas, including their arrangement structure.

In order to explain the idea of a GAN, we introduced the dialectical way and adapted each part of the proposed frame to fit with this logical structure. The proposed method is combining the loss functions of Spatial Gram and WGAN-GP methods in order to fulfil our requirements. The results of the translation show promising capabilities, especially for urban areas. The networks learn an adaptive loss from image pairs at hand, and regularized by the prescribed image style, which make it applicable for the task of SAR image translation.

As future works, we plan to go into deeper mathematic details and explanations of the Dialectical GAN. The combination of radar signal theory and deep learning needs to be investigated in order to describe the change of the basic unit (*e.g.*, point spread function). In addition, this paper is limited to the application of SAR images translations, now we are trying to understand the translation of SAR and optical images. In future, we would like to apply our techniques to other target areas and other sensors.

**Acknowledgments:** We thank the TerraSAR-X Science Service System for the provision of images (Proposal MTH-1118) and China Scholarship Council (Grant No. 201606030108)

**Author Contributions:** "Dongyang Ao and Mihai Datcu conceived and designed the experiments; Dongyang Ao performed the experiments; Dongyang Ao, Dumitru Octavian, Gottfried Schwarz and Mihai Datcu analyzed the data; Dumitru Octavian contributed data materials; Dongyang Ao proposed the method and wrote the paper."

**Conflicts of Interest:** The authors declare no conflict of interest.